\documentclass[onecolumn,showpacs,12pt]{revtex4}
\usepackage{bm,amsmath,amssymb} 
\usepackage{graphicx}
\usepackage{amssymb}
\usepackage{epsfig}
\usepackage[english]{babel}
\usepackage{latexsym}
\usepackage{graphics}
\usepackage{epsfig}
\usepackage{graphicx}
\usepackage{dcolumn}
\usepackage{amsmath}

\begin{document}


\title{Universality of the dynamic characteristic relationship of electron correlation in the two-photon double ionization process of a helium-like system}


\author{Fei Li$^{1,6}$, Yu-Jun Yang$^{2}$, Jing Chen$^{3,4}$, Xiao-Jun Liu$^{5}$, Zhi-Yi Wei$^{1,6}$ and Bing-Bing Wang$^{1,6}$}
\email{zywei@iphy.ac.cn; wbb@aphy.iphy.ac.cn}

\affiliation{$^1$Laboratory of Optical Physics, Beijing National Laboratory for Condensed Matter Physics, Institute of Physics, Chinese Academy of Sciences, Beijing 100190, China}
\affiliation{$^2$Institute of Atomic and Molecular Physics, Jilin University, Changchun 130012, China}
\affiliation{$^3$Institute of Applied Physics and Computational Mathematics, P. O. Box 8009, Beijing 100088, China}
\affiliation{$^4$HEDPS, Center for Applied Physics and Technology, Peking University, Beijing 100084, China}
\affiliation{$^5$State Key Laboratory of Magnetic Resonance and Atomic and Molecular Physics, Wuhan Institute of Physics and Mathematics, Innovation Academy for Precision Measurement Science and Technology, Chinese Academy of Science, Wuhan 430071, China}

\affiliation{$^6$University of Chinese Academy of Sciences, Beijing 100049,  China}
\date{\today}

\begin{abstract}
Universality of the dynamic characteristic relationship between the characteristic time $t_c$ and the two-electron Coulomb interaction energy $\overline{V}_{12}$ of the ground state in the two-photon double ionization process is investigated through changing the parameters of the two-electron atomic system and the corresponding laser conditions. The numerical results show that the product $t_{c}\overline{V}_{12}$ keeps constant around 4.1 in the cases of changing the nucleus charge, the electron charge, the electron mass, and changing simultaneously the nucleus charge and the electron charge. These results demonstrate that the dynamic characteristic relationship in the two-photon double ionization process is universal. This work sheds more light on the the dynamic characteristic relationship in ultrafast processes and may find its application in the measurement of attosecond pulses.
\end{abstract}

 \maketitle



Electron correlation plays an important role in the dynamic processes for a many-body system\cite{1,2,3,4,5,6,7,8,9,10}, where the time-dependent characteristic of the electron correlation is hard to be identified because of its attosecond timescale. Until recently, with the development of ultrafast laser technology and attosecond high-harmonic pulses, the investigation of the electron correlation in ultrafast dynamic processes becomes accessible in experiments. For example, the relative photo-emission timing is measured by pump-probe experiments\cite{6}, the electron correlation effects in the ionization processes of D$_2$ molecule has been uncovered successfully through combining ultrafast and synchrotron XUV sources with electron-ion 3D coincidence imaging techniques\cite{8}. These experimental techniques may be applied to strictly test some fundamental theoretical predictions.

Helium-like system is one of the simplest few-body system, which provides a benchmark system for the study of the time characteristics of the electron correlation effects in many-body dynamic processes. As mentioned by Hu\cite{7}, the electron correlation for the ground state of the helium-like atoms becomes stronger with the increase of the nuclear charge. Therefore, one expects that the dynamic process will also change for different  correlated electron systems. Hence, an interesting question is: Is there any general characteristics of the electron correlation which may be hold even if the parameters of the system changes? In this work, we demonstrate that the dynamic characteristic  relationship between the electron correlation and the characteristic time of the two-photon double ionization (TPDI) process of two-electron system is valid for various helium-like systems, e.g., Li$^+$ and Be$^{2+}$ ions. These results consolidate the universality of the dynamic characteristic relationship obtained in our previous work\cite{11}, which can be regarded as a general dynamic characteristic relationship in the atomic ionization processes.

We investigate the TPDI of the helium-like ion system by solving numerically the time-dependent schr\"{o}dinger equation (TDSE). In general, the field-free three-particle Schr\"{o}dinger equation can be written as
\begin{equation}
\label{equation1}
\begin{aligned}
  &\bigg(-\frac{\hbar^{2}}{2M}\nabla^{2}_{R_{0}}-\frac{\hbar^{2}}{2m}\nabla^{2}_{R_{1}}- \frac{\hbar^{2}}{2m}\nabla^{2}_{R_{2}}-\frac{Ze^{2}}{|\textbf{R}_{1}-\textbf{R}_{0}|}- \frac{Ze^{2}}{|\textbf{R}_{2}-\textbf{R}_{0}|}+\frac{e^{2}}{|\textbf{R}_{1}-\textbf{R}_{2}|}\bigg)\times\\ &\;\;\Psi (\textbf{R}_{0},\textbf{R}_{1},\textbf{R}_{2})=E_{NR}\Psi (\textbf{R}_{0},\textbf{R}_{1},\textbf{R}_{2}),
\end{aligned}
\end{equation}
where $M$ is the nucleus mass, $m$ is the electron mass, $\hbar$ is the reduced Planck constant, $Z$ is the atomic number, $e$ is the elementary charge, $\textbf{R}_{0}$ is the position vector of the nucleus, $\textbf{R}_{1}$ and $\textbf{R}_{2}$ are the position vectors of the two electrons, respectively, and  $E_{NR}$ is the non-relativistic energy. On making the standard transformation to scaled center-of-mass coordinate system\cite{12}, equation (\ref{equation1}) is reduced to the dimensionless form
\begin{equation}
\label{equation2}
  \begin{aligned}
  H_{0}\Psi_{n} (\textbf{r}_{1},\textbf{r}_{2})=E_{n}\Psi_{n} (\textbf{r}_{1},\textbf{r}_{2}),
   \end{aligned}
\end{equation} 
with 
\begin{equation}
\label{equation3}
  \begin{aligned} H_{0}=\bigg[-\frac{1}{2\gamma}\nabla^{2}_{r_{1}}-\frac{1}{2\gamma}\nabla^{2}_{r_{2}}-\frac{Z\xi}{r_{1}}-\frac{Z\xi}{r_{2}}+
  \frac{\xi^2}{|\textbf{r}_{1}-\textbf{r}_{2}|}\bigg],
  \end{aligned}
\end{equation} 
where $\gamma$ and $\xi$ are the parameters of the helium-like system, with (1) $\gamma=1, \xi=1$ for the case that only the nucleus charge $Ze$ changes, including $Z=3$ for Li$^+$ ion and $Z=4$ for Be$^{2+}$ ion; (2) $\gamma=1, \xi=b$ for the case that the electron charge changes from $-e$ into $-be$; (3) $\gamma=n, \xi=1$ for the case that the mass of the electron changes from $m$ into $nm$. Then the eigenenergy $E_{n}$ can be expressed as $E_{n}=E_{NR}/E_{s}$ in the case of ignoring the movement of the center-of mass, where, in the dimensionless process, $E_{s}=e^{2}/a_{\mu}$ is the unit of energy and $a_{\mu}=\hbar^{2}/\mu e^{2}$ is the unit of length with the reduced mass $\mu=Mm/(M+m)\approx m$. The field-free Schr\"{o}dinger equation (\ref{equation2}) can be solved through the variational method, and $\Phi(\textbf{r}_{1},\textbf{r}_{2})$ can be obtained by selecting an appropriate trial wave function which can be expressed by B-spline functions as
\begin{equation}
\label{equation4}
  \begin{aligned}
  \Psi_{n}(\textbf{r}_{1},\textbf{r}_{2})&=\sum_{i_{\alpha}, i_{\beta}, l_{\alpha}, l_{\beta} }C_{i_{\alpha},i_{\beta},l_{\alpha},l_{\beta}}[1+(-1)^{S}P_{12}]
   B_{i_{\alpha}}^{k}(r_{1})B_{i_{\beta}}^{k}(r_{2})\times\\
   &\hspace{5mm}\sum_{m_{\alpha}m_{\beta}}\langle l_{\alpha}m_{\alpha}l_{\beta}m_{\beta}|LM \rangle Y_{l_{\alpha}}^{m_{\alpha}}(\hat{\textbf{r}}_{1})Y_{l_{\beta}}^{m_{\beta}} (\hat{\textbf{r}}_{2}),
  \end{aligned}
\end{equation}
where $P_{12}$ is the permutation operator between electrons 1 and 2, $B_{i_{\alpha}}^{k}(r_{1})$ and $B_{i_{\beta}}^{k}(r_{2})$ are two B-spline functions of order $k$\cite{13,14}, $\langle l_{\alpha}m_{\alpha}l_{\beta}m_{\beta}|LM \rangle$ is the Clebsch-Gordan coefficient, $Y_{l_{\alpha}}^{m_{\alpha}}(\hat{\textbf{r}}_{1})$ and $Y_{l_{\beta}}^{m_{\beta}}(\hat{\textbf{r}}_{2})$ are the two spherical harmonics, and $S$, $L$, and $M$ are, respectively, the total spin of the two electrons, the total orbital angular momentum and its $z$-component.

Once the field-free $H_{0}$ eigenfunction $\Phi_{n}(\textbf{r}_{1},\textbf{r}_{2})$ is determined, the interaction between the helium-like system and ultrashort laser pulse can be solved. In the dipole approximation and the gauge transformation\cite{15}, the TDSE reduces to
\begin{equation}
\begin{aligned}
\label{equation5}
i\hbar\frac{\partial}{\partial t}\Phi(\textbf{r}_{1},\textbf{r}_{2},t)=[H_{0}+H_{int}(t)]\Phi(\textbf{r}_{1},\textbf{r}_{2},t),
\end{aligned}
\end{equation}
where
\begin{equation}
\begin{aligned}
\label{equation6}
H_{int}(t)=\xi e\textbf{E}(t)\cdot(\textbf{r}_{1}+\textbf{r}_{2}),
\end{aligned}
\end{equation}
and $\textbf{E}(t)$ is the electric field of laser pulse. In our simulation, the vector potential of laser pulse  is expressed as
\begin{equation}
\label{equation7}
  \begin{aligned}
    \textbf{A}(t)=-A_{0}e^{-(2\ln2)(t-t_{c})^2/\tau^{2}}\sin(\omega t)\hat{\textbf{e}}_{z},
  \end{aligned}
\end{equation}
and the corresponding electric field can be expressed as
\begin{equation}
\label{equation8}
  \begin{aligned}
    \textbf{E}(t)&=E_{0}e^{-(2\ln2)(t-t_{c})^2/\tau^{2}}\cos(\omega t)\hat{\textbf{e}}_{z}-(4\ln2)E_{0}(t-t_{c})\times\\
    &\hspace{5mm}e^{-(2\ln2)(t-t_{c})^2/\tau^{2}}\sin(\omega t)/(\omega\tau^{2})\hat{\textbf{e}}_{z},
  \end{aligned}
\end{equation}
where $E_{0}=A_{0}\omega$ is the electric-field amplitude, $t_{c}$ is the position of the laser pulse center on the time-axis, $\tau$ is the full width at half maximum (FWHM), $\omega$ is the center frequency of the laser pulse, and $\hat{\textbf{e}}_{z}$ is the unit vector of the polarization direction of the laser pulse. Specially, when the electron charge changes from $-e$ into $-be$, the equation(6) becomes
\begin{equation}
\begin{aligned}
\label{equation9}
H_{int}(t)=be\textbf{E}(t)\cdot(\textbf{r}_{1}+\textbf{r}_{2}).
\end{aligned}
\end{equation}
The time-dependent wave function $\Phi(\textbf{r}_{1},\textbf{r}_{2},t)$ can be expanded in terms of the field-free $H_{0}$ eigenfunctions $\Psi_{n}$. By substituting the expanded expression of the $\Phi(\textbf{r}_{1},\textbf{r}_{2},t)$ into equation (\ref{equation5}), one can obtain a set of coupled differential equations, which can be solved by the Adams method\cite{16}. Once the time-dependent wave function $\Psi(\textbf{r}_{1}, \textbf{r}_{2}, t)$ is determined, the probability distribution at the time $t_{f}$ for the two ionized electrons escaped with momenta $\textbf{k}_{1}$ and $\textbf{k}_{2}$ is obtained according to
\begin{equation}
\label{equation10}
  \begin{aligned}
    P(\textbf{k}_{1},\textbf{k}_{2})=|\langle\psi_{\textbf{k}_{1},\textbf{k}_{2}}(\textbf{r}_{1},\textbf{r}_{2})
  |\Phi(\textbf{r}_{1},\textbf{r}_{2},t_{f})\rangle|^{2},
  \end{aligned}
\end{equation}
where $\psi_{\textbf{k}_{1},\textbf{k}_{2}}(\textbf{r}_{1},\textbf{r}_{2})$ is the wave function of the uncorrelated double continuum state, which can be expressed as a symmetrized product of two independent-particle Coulomb wave functions\cite{17}. Therefore, the energy distribution of two ionized electrons can be expressed as
\begin{equation}
\label{equation11}
  \begin{aligned}
   P(E_{1},E_{2})=\iint k_{1}k_{2}P(\textbf{k}_{1},\textbf{k}_{2})d\hat{\textbf{k}}_{1}d\hat{\textbf{k}}_{2},
  \end{aligned}
\end{equation}
where $E_{1}=k_{1}^{2}/2$ and $E_{2}=k_{2}^{2}/2$ are the energies of the two ionized electrons. The change of Coulomb wave function caused by changing the electron charge or the electron mass is equivalent to the change of the nucleus charge. However, the field-free two-electron wave function does not have such property. In other words, the changes of field-free two-electron wave function caused by changing the nucleus charge, the electron charge, or the electron mass are independent with each other.

The universality of the dynamic characteristic relationship in the TPDI process can ensure its broad applicability in various correlated systems. Thus various independent correlated systems can be selected as testing samples of the universality of the dynamic characteristic relationship. The correlated system consisting of a nucleus and two electrons has the inherent properties such as mass and charge etc. when the mass or charge of a particle is changed, the whole three-body correlated system is changed accordingly, and these changes are independent as mentioned above. Therefore, it is reasonable to explore the universality of the dynamic characteristic relationship in the TPDI process by changing these properties of the correlated systems separately.

We first investigate the TPDI process of helium, $\rm Li^{+}$ and $\rm Be^{2+}$ ion to further understand the characteristic time.  Fig. \ref{figure1} shows the energy distributions of two ionized electrons for helium ((a)-(c)), $\rm Li^{+}$ ((d)-(f)) and $\rm Be^{2+}$ ion ((g)-(i)), where the energy distribution in the TPDI process changes from one peak into two peaks as the laser pulse duration increases. Hence, we may define a ratio $\eta=P_{mid}/P_{max}$, where $P_{max}$ is the maximum value of the ionization probability in energy space with its coordinates($E_{1}(P_{max})$, $E_{2}(P_{max})$) and $P_{mid}$ is the probability at the intersection of lines $E_1+E_2=E_t$ and $E_1=E_2$, where $E_1$ and $E_2$ are the energies of the two electrons and $E_t=E_1(P_{max})+E_2(P_{max})$. Fig. \ref{figure2} presents the ratio $\eta$\cite{11} as a function of pulse duration with the laser intensity of $1\times 10^{14}\rm W/cm^{2}$ for helium, $\rm Li^{+}$ and $\rm Be^{2+}$ ions. Based on our previous work\cite{11}, we also define the characteristic time as the pulse duration at the turning point of the ratio $\eta$ from $\eta=1$ to $\eta<1$ in Fig. \ref{figure2},  which corresponds to that the energy distribution changes from one peak to two peaks. More specifically, as shown in Fig.~\ref{figure2}, the characteristic time is about 105 asec, 65 asec and 46 asec for helium, $\rm Li^{+}$ and $\rm Be^{2+}$ ion, respectively. When the pulse duration is less than the characteristic time, the ratio $\eta$ equals to 1, which indicates that two ionized electrons carry mainly equal energy. In contrast, when the pulse duration is greater than the characteristic time, the ratio $\eta$ is less than 1 and gradually tends to a constant with the increase of the pulse duration, which means that two ionized electrons carry mainly unequal energy. In order to understand the phenomenon that this ratio tends to a constant,  we define the energy difference $\Delta E$ for two ionized electrons that possess the maximum ionization probability in the TPDI process. Fig. \ref{figure3} shows the energy difference $\Delta E$ as a function of pulse duration with the laser intensity of $1\times 10^{14}\rm W/cm^{2}$ for helium, $\rm Li^{+}$ and $\rm Be^{2+}$ ion. By comparing figures \ref{figure2} and \ref{figure3}, we may find that, as the pulse duration increases further, the energy difference $\Delta E$ tends to be the energy difference of $|I_{p2}-I_{p1}|$ when the ratio $\eta$ tends to a constant, where $I_{p1}$ and $I_{p2}$ are the first and second ionization energies of the atom or the ions, respectively. It is known that, as a sequential TPDI process, the energies carried by two ionized electrons are $E_{1}=\omega- I_{p1}$ and $E_{2}=\omega-I_{p2}$, respectively. Hence the energy difference of two ionized electron $\Delta E= E_1-E_2=|I_{p2}-I_{p1}|$ can be regard as a sign of the sequence TPDI in turn based on previous works\cite{1,3,18}.  Therefore,  the pulse duration for the ratio tending to a constant can be regarded as the intrinsic maximum time delay\cite{3} between the two ionization events, which can lead to a specific combination of final energies of the ejected electrons, i.e. $E_{1}=\omega- I_{p1}$ and $E_{2}=\omega-I_{p2}$. Moreover, it should be emphasized that the characteristic time we defined here is different from the maximum time delay which is defined in reference \cite{3} and our characteristic time is the minimum duration of the laser pulse where the ejected electrons may carry different energies at the end of the laser pulse, rather than the pulse duration which identifies the nonsequential ionization mechanisms. In addition,  we find that the ratio $\eta$ curve and energy difference $\Delta E$ curve keep unchanged as the laser intensity decreases from $1\times 10^{14}\rm W/cm^{-2}$ to $1\times 10^{13}\rm W/cm^{-2}$ and $1\times 10^{12}\rm W/cm^{-2}$.

We then investigate the change of the characteristic time by changing the charges of the nucleus and the electron, separately. Table \ref{table1} presents the corresponding parameters of the helium-like system as the nucleus charge increases from $1.5e$ to $4e$. The ground-state energy $E_{1S}$, the second ionization energy $I_{p2}$ and the center frequency of the laser pulse are presented in table \ref{table1}, where the center frequency of the laser pulse is chosen greater than the second ionization energy and less than the ground-state energy. As shown in table \ref{table1}, the Coulomb interaction energy of the ground state increases from 0.636 a.u. to 2.193 a.u., and the corresponding characteristic time of the TPDI process decreases from 6.529 a.u. to 1.901 a.u., i.e. from 158 asec to 46 asec. These results can be understood directly as follows: As the charge of the nucleus increases, the two electrons are closer to each other, hence the Coulomb interaction energy increases, and the electron correlation becomes stronger. On the other hand, the characteristic time decreases drastically with the increase of the nucleus charge. The interesting result is that the product of the Coulomb interaction energy of the ground state and the  characteristic time of the TPDI process for the systems with different nucleus charge almost keeps constant around 4.1.

The change of the charge and mass of electron to test universality of the dynamic characteristic relationship  can be understood from the perspective of the equivalent quality and charge\cite{19,20}, although they are fundamental physics constants. Table \ref{table2} presents the values of the characteristic time and the Coulomb interaction energy of the ground state for the case that the electron charge is changed from $-0.6e$ to $-1.3e$. It shows that the Coulomb interaction energy of the ground state increases from 0.230 a.u. to 1.884 a.u. with the increase of the absolute value of the electron charge, and the corresponding characteristic time of the TPDI process decreases from 18.347 a.u. to 2.190 a.u., i.e. from 444 asec to 53 asec. However, it is found that the product of the Coulomb interaction energy of the ground state and the  characteristic time of the TPDI process also keeps constant around 4.1. Table \ref{table3} shows the values of the characteristic time and the Coulomb interaction energy as the nucleus charge increases from $1.4e$ to $2.4e$ and the electron charge changes from $-0.7e$ to $-1.2e$ simultaneously. Table \ref{table3} illustrates that although the Coulomb interaction energy of the ground state increase from 0.227 a.u. to 1.964 a.u. and the characteristic time of the TPDI process decrease from 18.058 a.u to 2.066 a.u., i.e. from 437 asec to 50 asec, the product of the Coulomb interaction energy of the ground state and the characteristic time of the TPDI process once again almost keeps constant around 4.1.

We finally study that how does the product of the characteristic time of the TPDI process and the Coulomb interaction energy of the ground state change when the electron mass are changed. Table \ref{table4} shows that as the electron mass increases from $0.4m$ to $1.6m$, the Coulomb interaction energy of the ground state increases from 0.379 a.u. to 1.515 a.u., the characteristic time of the TPDI process decreases from 10.826 a.u. to 2.686 a.u., i.e. from 262 asec to 65 asec,  and also the product of the Coulomb interaction energy of the ground state and the characteristic time of the TPDI process almost keeps constant around 4.1.

Figure \ref{figure4} shows that the characteristic time of the TPDI process varies with the Coulomb interaction energy of the ground state for the cases: (1) the change of the nucleus charge, (2) the change of the electron charge, (3) the simultaneous change of the nucleus and the electron charge, (4) the change of the electron mass. We found that the characteristic time $t_{c}$ of the TPDI process is inversely proportional to the Coulomb interaction energy $\overline{V}_{12}$ of the ground state, i.e. there is a dynamic characteristic relationship $t_{c}\overline{V}_{12}\approx 4.1$ in the TPDI process. We can also see that the dynamic characteristic relationship is true in the correlated systems mentioned above from another perspective, in other words, the dynamic characteristic relationship in the TPDI process is universal.

In conclusion, we have investigated the TPDI process of the various helium-like three-body correlated system by solving numerically time-dependent schr\"{o}dinger equation. we found that the dynamic characteristic relationship $t_{c}\overline{V}_{12}\approx 4.1$ in the TPDI process is universal. We also illustrate that the characteristic time defined here is clearly different from the time delay in reference \cite{3}, and the dynamic relationship here may help us to understand the ultrafast TPDI processes more deeply.

This work was supported by the National Natural Science Foundation of China under Grant Nos. 91850209, 11774129 and 11774411, and the National Key Research and
Development Program (No. 2019YFA0307700, and No. 2016YFA0401100). We thank all the members of SFAMP club for helpful discussions.

\newpage{

\begin{table}[!tb]
\centering
\setlength{\abovecaptionskip}{10pt}
\setlength{\belowcaptionskip}{10pt}
\caption{\linespread{1.5}\selectfont The parameters for the case of the change of the nucleus charge. The nucleus charge $Q_{c}$, the ground state energy $E_{1S}$, the two-electron Coulomb interaction energy $\overline{V}_{12}$, the second ionization energy $I_{p_2}$, the center  frequency $\omega$, and the characteristic time $t_{c}$ obtained by the time-dependent Schr\"{o}dinger equation.}
\label{table1}
\begin{tabular}{ccccccccc}
  \hline
  \hline
  $Q_{c}$ & $E_{1S} (a.u.)$ & $\overline{V}_{12} (a.u.)$ & $I_{p2}$ & $\omega (a.u.)$ & $t_{c} (asec)$ & $t_{c} (a.u.)$ & $\overline{V}_{12}\times t_{c}$\\
  \hline
  $1.5e$    & -1.465  & $0.636$ & $1.125$ & $1.3$ & $158$ & $6.529$ & $4.152$ \\
  $1.732e$  & -2.070  & $0.780$ & $1.5$ & $1.8$ & $127$ & $5.248$ & $4.093$ \\
  $2e$      & -2.903  & $0.947$ & $2$ & $2.4$ & $105$ & $4.339$ & $4.109$ \\
  $2.236e$  & -3.755  & $1.094$ & $2.5$ & $3$ & $91$  & $3.760$ & $4.114$ \\
  $2.45e$   & -4.624  & $1.227$ & $3$ & $3.4$ & $81$  & $3.347$ & $4.107$ \\
  $2.5e$    & -4.811  & $1.258$ & $3.125$ & $3.6$ & $79$  & $3.264$ & $4.107$ \\
  $2.8e$    & -6.244  & $1.445$ & $3.92$ & $4.7$ & $70$  & $2.893$ & $4.180$ \\
  $3e$      & -7.28   & $1.568$ & $4.5$ & $5$ & $65$  & $2.686$ & $4.212$ \\
  $4e$      & -13.654 & $2.193$ & $8$ & $9.5$ & $46$  & $1.901$ & $4.196$ \\
  \hline
  \hline
\end{tabular}
\end{table}

\begin{table}[!tb]
\centering
\setlength{\abovecaptionskip}{10pt}
\setlength{\belowcaptionskip}{10pt}
\caption{\linespread{1.5}\selectfont The parameters for the case of the change of the electron charge. The electron charge $Q_{e}$, the ground state energy $E_{1S}$, the two-electron Coulomb interaction energy $\overline{V}_{12}$, the second ionization energy $I_{p_2}$, the center  frequency $\omega$, and the characteristic time $t_{c}$ obtained by the time-dependent Schr\"{o}dinger equation.}\label{table2}
\begin{tabular}{cccccccc}
  \hline
  \hline
  $Q_{e}$  & $E_{1S} (a.u.)$ & $\overline{V}_{12} (a.u.)$ & $I_{p2}$ & $\omega (a.u.)$ & $t_{c} (asec)$ & $t_{c} (a.u.)$ & $\overline{V}_{12}\times t_{c}$\\
  \hline
  $-0.6e$   & -1.190 & $0.230$ & $0.72$ & $0.8$ & $444$  & $18.347$ & $4.220$ \\
  $-0.7e$   & -1.568 & $0.355$ & $0.98$ & $1.2$ & $284$  & $11.736$ & $4.166$ \\
  $-0.8e$   & -1.983 & $0.515$ & $1.28$ & $1.8$ & $196$  & $8.099$ & $4.171$ \\
  $-0.9e$   & -2.429 & $0.712$ & $1.62$ & $2.1$ & $140$  & $5.785$ & $4.119$ \\
  $-e$      & -2.903 & $0.947$ & $2$ & $2.4$ & $105$ & $4.339$ & $4.109$ \\
  $-1.1e$   & -3.400 & $1.221$ & $2.42$ & $3$ & $81$ & $3.347$ & $4.087$ \\
  $-1.2e$   & -3.916 & $1.533$ & $2.88$ & $3.5$ & $64$ & $2.645$ & $4.054$ \\
  $-1.3e$   & -4.448 & $1.884$ & $3.38$ & $4$ & $53$ & $2.190$ & $4.126$ \\
  \hline
  \hline
\end{tabular}
\end{table}

\begin{table}[!tb]
\centering
\setlength{\abovecaptionskip}{10pt}
\setlength{\belowcaptionskip}{10pt}
\caption{\linespread{1.5}\selectfont The parameters for the case of the change of the nucleus charge and the electron charge. The nucleus charge $Q_{c}$, the electron charge $Q_{e}$, the ground state energy $E_{1S}$,  the two-electron Coulomb interaction energy $\overline{V}_{12}$, the second ionization energy $I_{p_2}$, the center frequency $\omega$, and the characteristic time $t_{c}$ obtained by the time-dependent Schr\"{o}dinger equation.}\label{table3}
\begin{tabular}{ccccccccc}
  \hline
  \hline
  $Q_{c}$ & $Q_{e}$  & $E_{1S} (a.u.)$ & $\overline{V}_{12} (a.u.)$ & $I_{p2}$ & $\omega (a.u.)$ & $t_{c} (asec)$ & $t_{c} (a.u.)$ & $\overline{V}_{12}\times t_{c}$\\
  \hline
  $1.4e$   & $-0.7e$   & -0.697 & $0.227$ & $0.480$ & $0.52$ & $437$  & $18.058$ & $4.099$ \\
  $1.6e$   & $-0.8e$   & -1.189 & $0.388$ & $0.819$ & $1.0$ & $256$  & $10.579$ & $4.104$ \\
  $1.8e$   & $-0.9e$   & -1.905 & $0.621$ & $1.312$ & $1.6$ & $159$  & $6.570$ & $4.080$ \\
  $2e$     & $-e$      & -2.903 & $0.947$ & $2$ & $2.4$ & $105$ & $4.339$ & $4.109$ \\
  $2.2e$   & $-1.1e$   & -4.250 & $1.386$ & $2.928$ & $3.6$ & $71$ & $2.934$ & $4.066$ \\
  $2.4e$   & $-1.2e$   & -6.019 & $1.964$ & $4.147$ & $5$ & $50$ & $2.066$ & $4.058$ \\
  \hline
  \hline
\end{tabular}
\end{table}

\begin{table}[!tb]
\centering
\setlength{\abovecaptionskip}{10pt}
\setlength{\belowcaptionskip}{10pt}
\caption{\linespread{1.5}\selectfont The parameters for the case of the change of the electron mass. The electron mass $m_{e}$, the ground state energy $E_{1S}$, the two-electron Coulomb interaction energy $\overline{V}_{12}$, the second ionization energy $I_{p_2}$, the center frequency $\omega$, and the characteristic time $t_{c}$ obtained by the time-dependent Schr\"{o}dinger equation.}\label{table4}
\begin{tabular}{cccccccc}
  \hline
  \hline
  $m_{e}$ & $E_{1S} (a.u.)$ & $\overline{V}_{12} (a.u.)$ & $I_{p2}$ & $\omega (a.u.)$ & $t_{c} (asec)$ & $t_{c} (a.u.)$ & $\overline{V}_{12}\times t_{c}$\\
  \hline
  $0.4m$   & -1.161 & $0.379$ & $0.8$ &  $0.95$ & $262$ & $10.826$ & $4.103$ \\
  $0.6m$   & -1.742 & $0.568$ & $1.2$ &  $1.3$ & $174$ & $7.190$ & $4.084$ \\
  $0.8m$   & -2.322 & $0.758$ & $1.6$ &  $1.7$ & $129$  & $5.331$ & $4.041$ \\
  $m$      & -2.903 & $0.947$ & $2$ &    $2.4$ & $105$  & $4.339$ & $4.109$ \\
  $1.2m$   & -3.483 & $1.136$ & $2.4$ &  $3$ & $86$  & $3.554$ & $4.037$ \\
  $1.4m$   & -4.064 & $1.326$ & $2.8$ &  $3.4$ & $75$  & $3.099$ & $4.110$ \\
  $1.6m$   & -4.644 & $1.515$ & $3.2$ &  $4.2$ & $65$  & $2.686$ & $4.069$ \\
  \hline
  \hline
\end{tabular}
\end{table}

\begin{figure}
\includegraphics[width=8.5 cm]{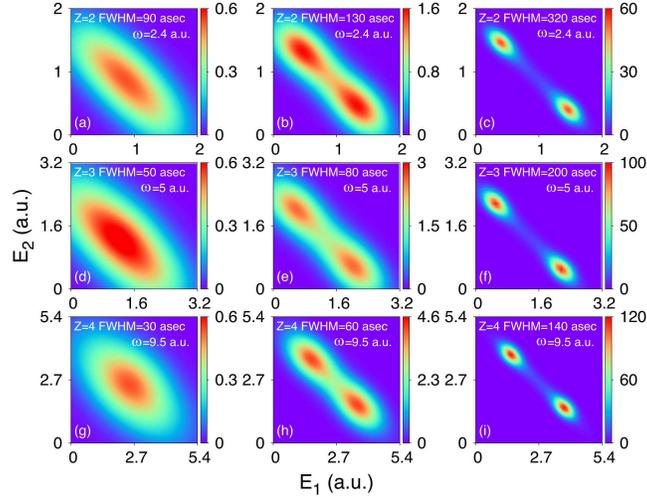}
\caption{Energy distribution of two escaped electrons of (a)-(c) helium atom, (d)-(f) $\rm Li^{+}$ ion, (g)-(i)$\rm Be^{2+}$ ion exposed to different laser pulse. The laser pulse has a Gaussian envelope around the peak intensity of $1\times10^{14}\,\rm W/cm^{2}$. The center photon energy is 2.4 a.u. for helium atom, 5.0 a.u. for $\rm Li^{+}$ ion and 9.5 a.u. for $\rm Be^{2+}$ ion. The FWHM is 90 asec, 130 asec, 320 asec  for helium atom, 50 asec, 80 asec, 200 asec  for $\rm Li^{+}$ ion, and 30 asec, 60 asec, 140 asec  for $\rm Li^{+}$ ion. The colors bars are in units of $10^{-6}$ for helium atom, $10^{-10}$ for $\rm Li^{+}$ ion and $\rm Be^{2+}$ ion.}
\label{figure1}
\end{figure}

\begin{figure}
\includegraphics[width=8.5 cm]{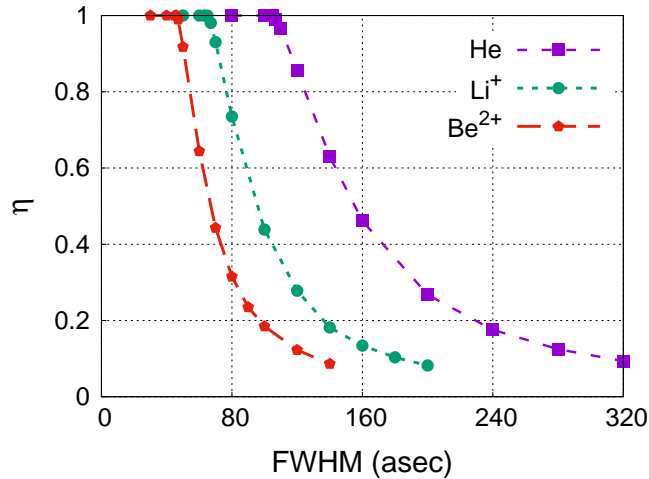}
\caption{The ratio $\eta$ varies with the FWHM of the laser pulse. The laser pulse has a Gaussian envelope around the peak intensity of $1\times10^{14}\,\rm W/cm^{2}$. The center photon energy is 2.4 a.u. for helium atom, 5.0 a.u. for $\rm Li^{+}$ ion and 9.5 a.u. for $\rm Be^{2+}$ ion.}
\label{figure2}
\end{figure}

\begin{figure}
\includegraphics[width=8.5 cm]{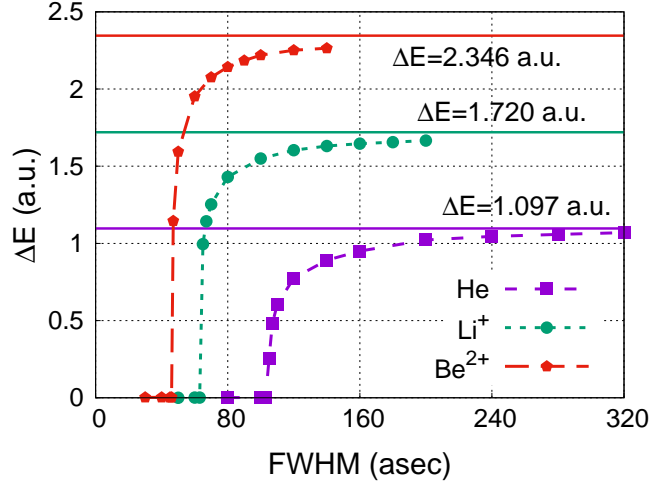}
\caption{The energy difference $\Delta E$ varies with the FWHM of the laser pulse. The laser pulse has a Gaussian envelope around the peak intensity of $1\times10^{14}\,\rm W/cm^{2}$. The center photon energy is 2.4 a.u. for helium atom, 5.0 a.u. for $\rm Li^{+}$ ion and 9.5 a.u. for $\rm Be^{2+}$ ion. The horizontal lines represent the energy difference $\Delta E=|I_{p2}-I_{p1}|$.}
\label{figure3}
\end{figure}

\begin{figure}
\includegraphics[width=8.5 cm]{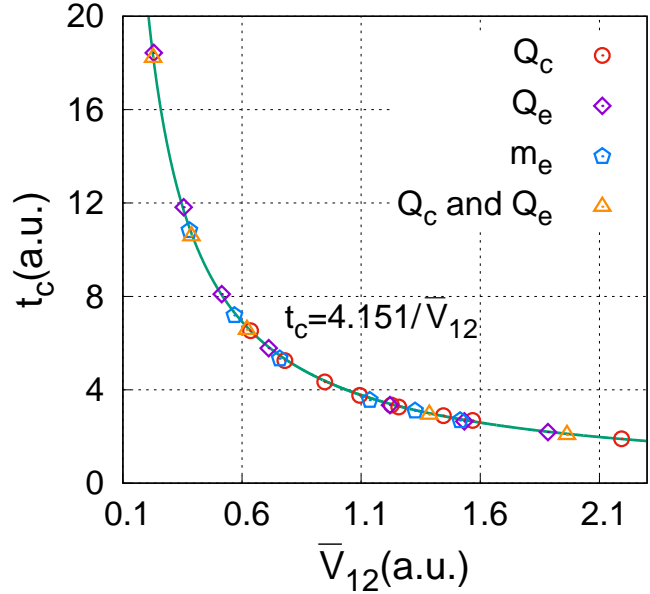}
\caption{The characteristic time of the TPDI varies with the Coulomb interaction energy for different cases: (1) the open circles represent the change of the nucleus charge; (2) the open diamonds represent the change of the electron charge; (3) the open pentagons represent the simultaneous change of the nucleus and electron charge; (4) the open triangles represent the change of the electron mass. The solid line is fitting curve for the data points. The laser pulse has a Gaussian envelope around the peak intensity of $1\times10^{14}\,\rm W/cm^{2}$.}
\label{figure4}
\end{figure}

\end{document}